\documentclass[aps,prl,twocolumn,groupedaddress,showpacs]{revtex4-1}

\usepackage{epsfig}
\usepackage{mathrsfs}
\usepackage{float}

\begin{document} 

\title{Nanometer-scale Exchange Interactions Between Spin Centers in Diamond} 
\author{V. R. Kortan}
\affiliation{Optical Science and Technology Center and Department of Physics and Astronomy, University of Iowa, Iowa City, IA 52242}
\author{C. \c{S}ahin}
\affiliation{Optical Science and Technology Center and Department of Physics and Astronomy, University of Iowa, Iowa City, IA 52242}
\author{M. E. Flatt\'e}
\affiliation{Optical Science and Technology Center and Department of Physics and Astronomy, University of Iowa, Iowa City, IA 52242}

\begin{abstract}
Exchange interactions between isolated pairs of spin centers in diamond have been calculated, based on an accurate atomistic electronic structure for diamond and any impurity atoms, for spin-center separations up to 2~nm. The exchange interactions exceed dipolar interactions for spin center separations less than 3~nm.   NV$^-$ spin centers, which are extended defects, interact very differently depending on the relative orientations of the symmetry axis of the spin center and the radius vector connecting the pair. Exchange interactions between transition-metal dopants behave similarly to those of NV$^-$ centers. The Mn\---Mn exchange interaction decays with a much longer length scale than the Cr\---Cr and Ni\---Ni exchange interactions, exceeding  dipolar interactions for Mn\---Mn separations less than 5~nm. Calculations of these highly anisotropic and spin-center-dependent interactions provide the potential for design of the spin-spin interactions for novel nanomagnetic structures.
\end{abstract}

\maketitle 

A single spin, such as from a defect or dopant, can control the properties of a nanomagnetic system\cite{Koenraad2011}, suggesting pathways to constructing novel magnetic materials or magnetic behavior through  designed assembly {\it e.g.} of  spins in metals, insulators, and semiconductors\cite{Heinrich2004,Kitchen2006,Toyli2010,Khajetoorians2010,Lee2010,Khajetoorians2011,Spinelli2014}. Spin centers in wide-gap semiconductors such as diamond exhibit exceptionally long room-temperature  spin coherence times\cite{Balasubramanian2009}, permitting coherent interactions among such spin centers over length scales of many nanometers, and the corresponding shaping of spin dynamics in the spin assemblies. As the interactions occur through weak, long-range, largely isotropic dipolar interactions\cite{Ohno2012,Myers2014} the interaction effects on spin dynamics are slow (less than 1~$\mu$eV).  Continued improvement of control in spin-center positioning, such as through ion implantation\cite{Rabeau2005, Aharonovich2010, Toyli2010}, will lead to assemblies with short-range coupling, where exchange interactions may dominate over dipolar interactions, producing anisotropic\cite{Kitchen2006} interactions that are orders of magnitude greater than dipolar interactions. The current focus on NV$^-$ centers in diamond, due especially to the convenience of its levels and optical selection rules for spin initialization and readout\cite{Jelezko2006}, may also shift to other spin centers that are easier to address and manipulate electrically, especially transition-metal dopants that possess partially-filled $d$ levels\cite{Chanier2012b,Chanier2012a}.

Here we construct a highly-accurate theoretical description of the spin center in bulk diamond, and a very efficient theoretical methodology to evaluate the exchange-coupling between spins in diamond, including both NV$^-$ centers and transition-metal spin centers. We include the weak spin-orbit interaction in bulk diamond and the strong spin-orbit interaction of a transition-metal dopant, as well as the dependence of an NV$^-$ spin center's interaction on the N-V axis direction. We find that exchange interactions dominate over dipolar interactions for spin-center separations smaller than 3~nm, except for the more delocalized Mn spins, which are exchange-dominated for separations less than 5~nm.  The theoretical techniques that have been previously applied to diamond find calculations of spin-spin interactions very challenging, either (as with density functional theory\cite{Weber2010,Chanier2012a,Chanier2012b}) due to the very large supercell sizes required for such calculations, or (as with symmetry-based group-theory analyses\cite{Dohery2012}) due to the inability to constrain the problem to a very small number of experimentally-determined quantities.  Our approach is a rigorously tested $spds^*$ description of the bulk electronic structure\cite{Jancu1998} and a set of effective impurity potentials, including for $d$ states, that replicate the energies of the spin-center states found in density functional theory calculations or experimental measurements. Once those are known the electronic properties of the pair are efficiently evaluated using a Green's function-based Koster-Slater method\cite{Koster1954} as described in Ref.~\cite{Tang2004}, and here extended to the $spds^*$ system required to accurately describe bulk diamond and the $d$ levels of transition-metal dopants. This approach\cite{Tang2004}, by exactly solving for the electron propagator in the regions between defects, permits calculations of the exchange interaction of a defect pair to proceed with a rapid speed that is independent of the defect separation.

The Hamiltonian for a point defect (impurity atom or vacancy) has the  form $H = H_0+V$, where $H_0$ is the $spds^*$ Hamiltonian of Ref.~\onlinecite{Jancu1998} and 
\begin{eqnarray}
V &=& \sum_{\ell,m,s}U^{os}_{\ell m s}c^\dagger_{ \ell m s}({\bf R_0})c_{\ell m s}({\bf R_0}) \nonumber\\
&&+ \sum_{j=1}^4\sum_\ell U^{nn}_{\ell m s}c^\dagger_{\ell  m s}({\bf R_j})c_{\ell m s}({\bf R_j})\\
&& + (2/3)\sum_{\ell,m,s}\Delta_{\ell}[ c^\dagger_{\ell m s} ({\bf R_0}) c_{\ell m+1 s-1}({\bf R_0})+ {\rm H. c.}].\nonumber
\end{eqnarray}
Here $U_{\ell m s}$ is the energy difference for the orbital with spin $s$, angular momentum $\ell$ and azimuthal quantum number $m$, either at the point defect site ($U^{os}$) or at the nearest neighbors ($U^{nn}$), and $\Delta_\ell$ is the point defect's spin-orbit interaction for the $\ell$ angular-momentum states. $c^\dagger_{\ell m 
  s}({\bf R})$ ($c_{\ell s}({\bf R})$) is the creation (annihilation) operator
for a spin-$s$ electron in the $\ell$, $m$ orbital at site ${\bf R}$. The point defect
is located at ${\bf R_0}$, and the four nearest-neighbor sites are
labeled by ${\bf R_1}$-${\bf R_4}$.  The spin-orbit potential has been calculated from atomic energies\cite{Moore1949, Moore1952, Moore1958} and using the Land\'e interval rule. Spin-orbit interactions are positive for angular-momentum shells less than half full, and negative otherwise. For transition-metal dopants, to position the $d$ states of correct tetrahedral symmetry ($t_2$ or $e$) at the correct locations within the diamond band gap, $U^{os}$ magnetic and nonmagnetic potentials are determined for the $t_2$ and $e$ states, and reported in Table~\ref{tmpotentials}. $U^{nn}=0$ for transition-metal dopants. For the NV$^-$ spin center, defect potentials are only required on the $p$ orbitals, however the shift in the atomic positions requires nonzero defect potentials on the nearest neighbors as well. These values are reported in Table~\ref{nvpotentials}. 

We calculate the retarded Green's function for the bulk Hamiltonian $H_0$,
$G_0({\bf k},\omega)=[\omega-
H_0({\bf k})+i\delta]^{-1}$, and from this the real-space Green's function  $G_0({\bf R_i},{\bf R_j},\omega)$, where $G_0$ is a matrix with rows and columns labeled by $\ell$, $m$, and $s$. The properties of the defects, either point defects or pairs, are determined from solving the Dyson equation in real space, 
\begin{equation}
 G(\omega)  =  \left[I- G_0(\omega) V\right]^{-1} G_0(\omega) \;.\label{dyson}
\end{equation}
Due to the limited number of positions in real space where the potential is non-zero, Eq.~(\ref{dyson}) can be solved rapidly once the $G_0({\bf R_i},{\bf R_j},\omega)$ have been tabulated.

\begin{table} \setlength{\tabcolsep}{8pt}
\caption{On-site potentials (eV) for transition-metal impurities in diamond, including the nonmagnetic and magnetic potentials for $d$ electrons of $t_2$ and $e$ symmetry, and the spin-orbit interaction strength $\Delta$ for $p$ and $d$ electrons, in eV.}
\begin{tabular}{l c c c c c c}
\hline\hline
&\multicolumn{2}{c}{nonmagnetic}&\multicolumn{2}{c}{magnetic}&\multicolumn{2}{c}{spin-orbit}\cr
&$t_2$&$e$&$t_2$&$e$&$p$&$d$\\[0.5ex]
\hline \\
Cr&-18.89&-21.45&-0.26&-1.85&0.09&0.02\\[1ex]
Mn&-19.30&-22.50&-0.14&-1.00&-0.03&-0.08\\[1ex]
Fe&-19.20&-23.15&0&0&-0.15&-0.12\\[1ex]
Co&-20.57&-24.64&-0.26&-0.21&-0.10&-0.19\\ [1ex]
Ni&-21.67&-27.03&-0.43&-0.38&-0.08&-0.33\\[1ex]
\hline\hline
\end{tabular}\label{tmpotentials}
\end{table}

\begin{table} \setlength{\tabcolsep}{6pt}
\caption{On-site and nearest-neighbor $p$-orbital potentials, magnetic and nonmagnetic, for nitrogen and a vacancy in diamond.}
\begin{tabular}{l c c c c}
\hline\hline
&\multicolumn{2}{c}{on-site}&\multicolumn{2}{c}{nearest-neighbor}\cr
&nonmagnetic&magnetic&nonmagnetic&magnetic\\[0.5ex]
\hline \\
N&-5.33&2.93&0&0\\ [1ex]
V&50&0&-0.26&-2.97\\[1ex]
\hline\hline
\end{tabular}\label{nvpotentials}
\end{table}

\begin{figure}
\includegraphics[width=\columnwidth]{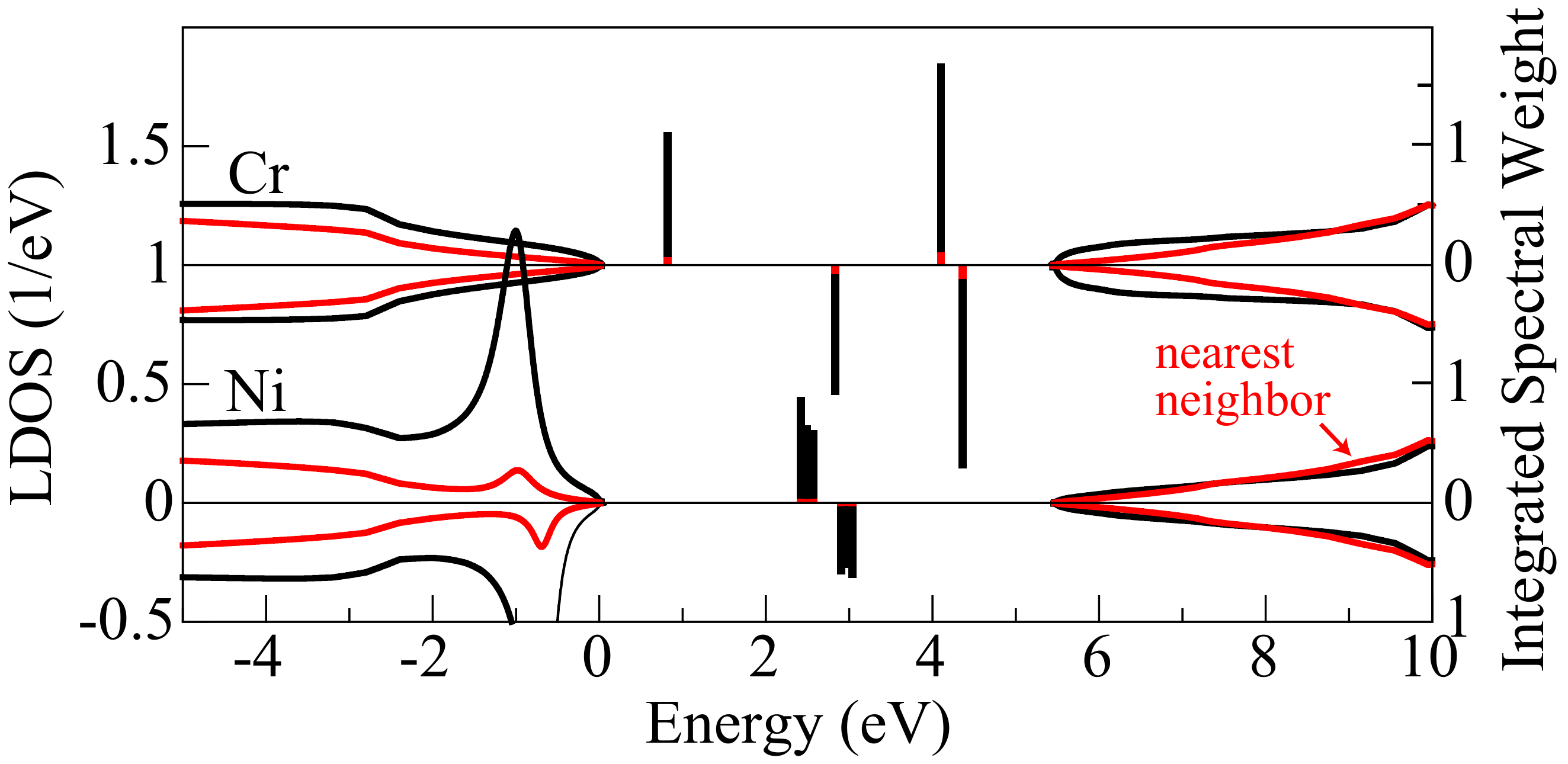}
\caption{Spin resolved local density of states (LDOS) on the impurity site and nearest neighbor carbon site for Cr and Ni spin centers.  The continuum states in the conduction and valance bands are plotted on the scale of the left axis.   The probabilities of finding the electron on the impurity for mid-gap impurity states are plotted on the scale of the left axis.  The nearest-neighbor contributions are  in red, whereas the on-site contributions are in black. }
\label{ldos}
\end{figure} 

Figure~\ref{ldos} compares the on site and nearest neighbor spin resolved local density of states (LDOS) for the two transition-metal spin-$1$ dopants, Ni and Cr.  Within the diamond band gap, the Cr spin center forms one doubly-degenerate spin-up and one doubly-degenerate spin-down $e$ level as well as one  triply degenerate spin-up and one triply-degenerate spin-down $t_{2}$ level.  The ground state for Cr has two electrons in the  spin-up $e$ state and the rest empty.  The Ni dopant levels are arranged differently, with the $t_{2}$ levels  in the gap and the $e$ levels below the edge of the valence band, showing as a broad resonance.  The $t_{2}$ levels for Ni show a visible splitting in Fig.~\ref{ldos} due to the large spin orbit coupling for Ni.  The ground state for the Ni spin center has two electrons in the spin-up $t_{2}$ states.  As found in Ref.~\onlinecite{Chanier2012b,Chanier2012a} with density functional theory calculations, the  Cr ground state possesses more spectral weight  on the site of the  dopant than the Ni ground state, with a ratio of $\sim$~2:1. The construction of the NV$^-$ center requires tracking different mid-gap levels. The NV$^-$ center exhibits four levels in the gap, the lower two have $a_1$ symmetry and the upper two are  spin-split, orbitally-degenerate $e_x$ and $e_y$ levels, all of which originate from $p$ orbitals ($t_{2}$ character)\cite{Weber2010}.  The ground state for the NV$^-$ center fills electrons up through the spin-up $e_x$ and $e_y$ states.

\begin{figure}
\includegraphics[width=\columnwidth]{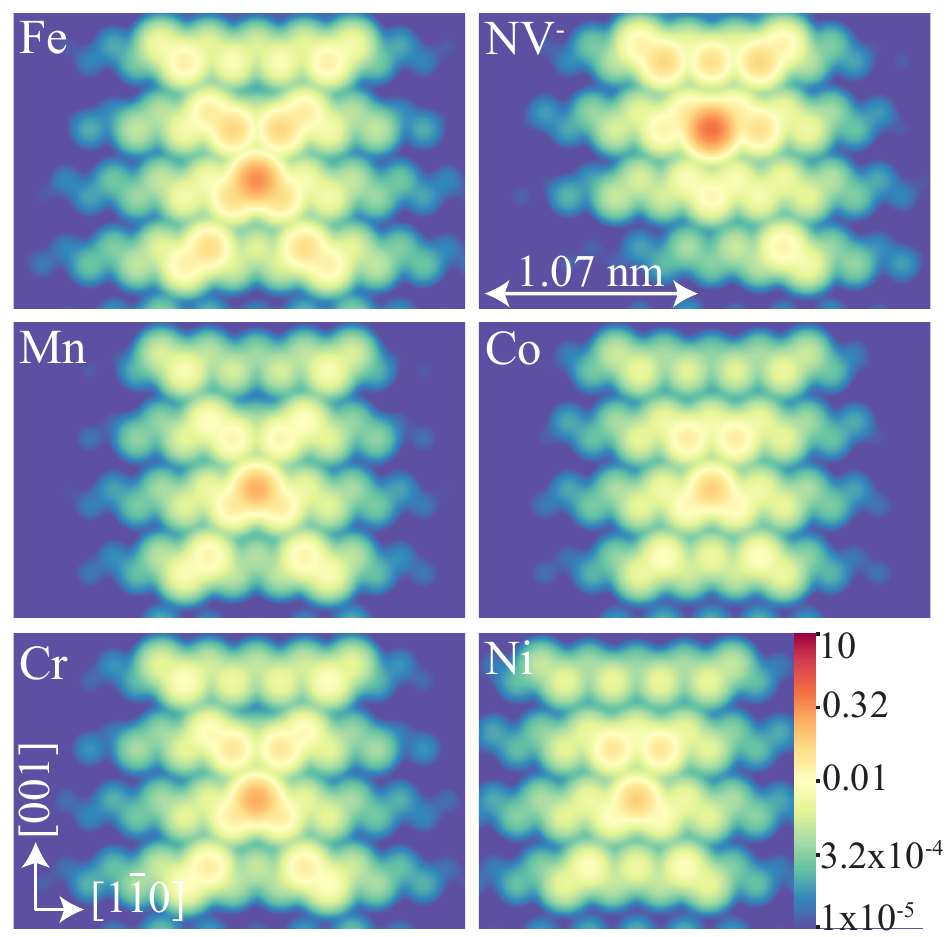}
\caption{Real space probability density for (a) Fe, (b) NV$^-$, (c) Mn, (d) Co, (e) Cr, (f) Ni dopants with any background contribution from the homogeneous diamond crystal removed.  The slices are taken in the (110) plane and three atomic layers above the dopant.  The logarithmic color scale for all plots is the same, and is in units of the inverse volume of an atomic site.}
\label{wavefunction}
\end{figure}

These trends are reflected in the real space probability density of the highest occupied molecular orbital (HOMO) of each of the spin centers in Fig.~\ref{wavefunction}.  The ground state spins for each dopant in diamond are Fe: spin 0, Mn and Co: spin 1/2, and NV$^-$, Cr and Ni: spin 1.  All of the transition-metal dopant HOMOs show the same overall spatial symmetry regardless of spin, which is expected because the propagation of electron waves in the host material most determines the probability density symmetry\cite{Tang2004}.  The Fe, Mn, and Cr dopants all have $e$-like HOMOs whereas the NV$^-$, Co and Ni spin centers have $t_{2}$-like HOMOs, and therefore among the point defects Fe, Mn and Cr all have larger wave function probability near the dopant location and appear less extended than the Co and Ni wave functions.

\begin{figure}[t]
\includegraphics[width=\columnwidth]{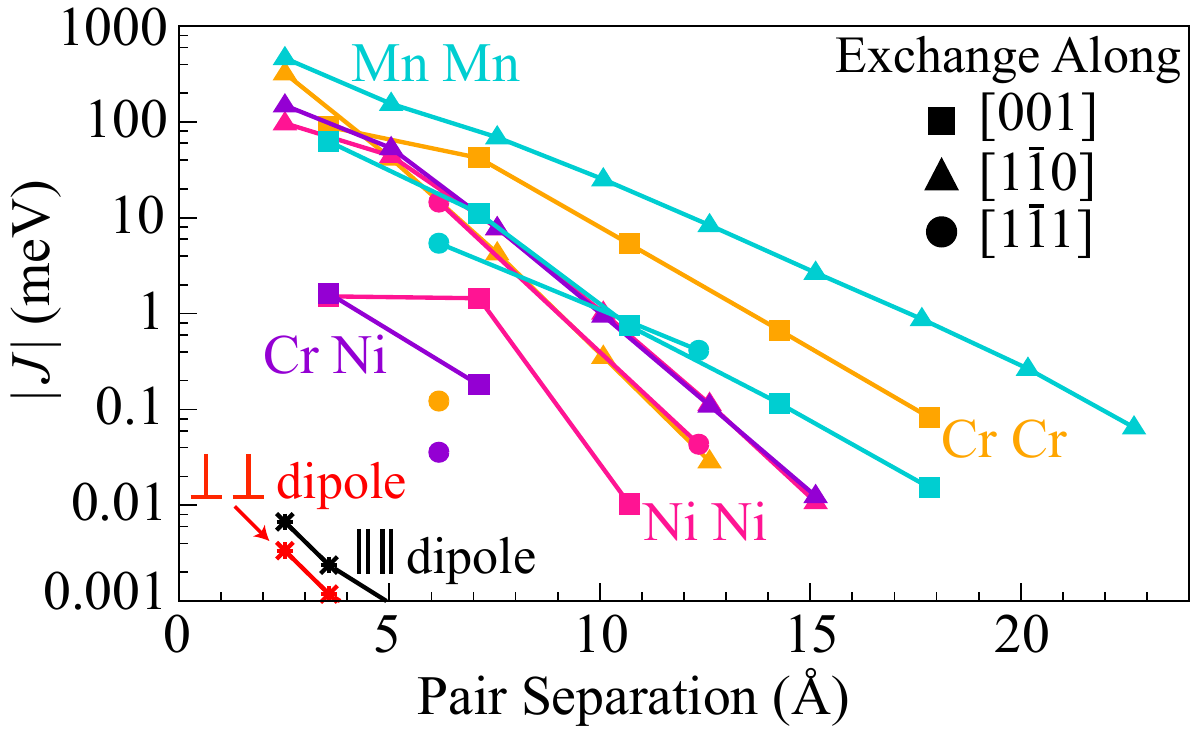}
\caption{Magnitude of the exchange interaction for several parings of transition metal spin centers along $[001]$, $[1\bar10]$ and $[1\bar11]$ denoted by triangles, squares and circles respectively.  The four sets of spin center pairs are Mn-Mn (light blue), Ni-Ni (pink), Cr-Cr (gold) and Cr-Ni (purple).}
\label{exchange}
\end{figure}

Once the properties of an individual spin center have been determined the exchange interaction between two can be calculated by comparing the energies of filled mid-gap states for parallel and antiparallel alignment of the spin centers\cite{Tang2004,Kitchen2006}. The exchange interaction found between pairs of transition metal spin centers is shown in Fig.~\ref{exchange}.  For pairs spaced along the $[1\bar10]$ direction the Mn\--Mn pair has the largest and slowest-decaying exchange, followed by Cr\---Cr pairs and then  Ni\---Ni pairs. 
The exchange interaction between Cr and Ni appears often smaller than either the Cr\---Cr or Ni\---Ni exchange, which is likely due to the smaller hybridizations of the energy levels of Cr and Ni (relative to homodopant pairs) due to their different energies.  Along the $[001]$ direction the Ni\--Ni pair does not decrease logarithmically for the closest pair spacings.  The exchange interaction along the $[1\bar11]$ interaction is the largest for the Ni\---Ni pair and excluding the Ni\---Ni pair it is the direction for which the exchange interaction between other transition metal pairs is the least.  At pair spacings greater than $\sim 2$~nm the energy broadening of the calculation ($10$\ $\mu$eV) limits the ability to resolve the exchange splittings, and for several pairs of spin centers the exchange interaction is obscured at shorter distances by this broadening.  At the first nearest neighbor spacing in the $[001]$ direction and the first and second nearest neighbor spacing in the $[1\bar10]$ direction the energy broadening in the calculation is on the order of 1 meV and thus the error for these points is larger than the others.  The exchange interaction is strongly anisotropic and can vary greatly depending on the direction of interaction, the energy of the spin center states as well as the symmetry of the HOMO (which produces the greatest hybridization and splitting), ie $e$ or $t_2$. For all these calculations the strength of the exchange interaction exceeds the dipolar interaction (also shown on Fig.~\ref{exchange}) by orders of magnitude. Only for spin center separations in excess of 3~nm would the dipolar interaction become comparable to the exchange interaction.

\begin{figure}
\includegraphics[width=\columnwidth]{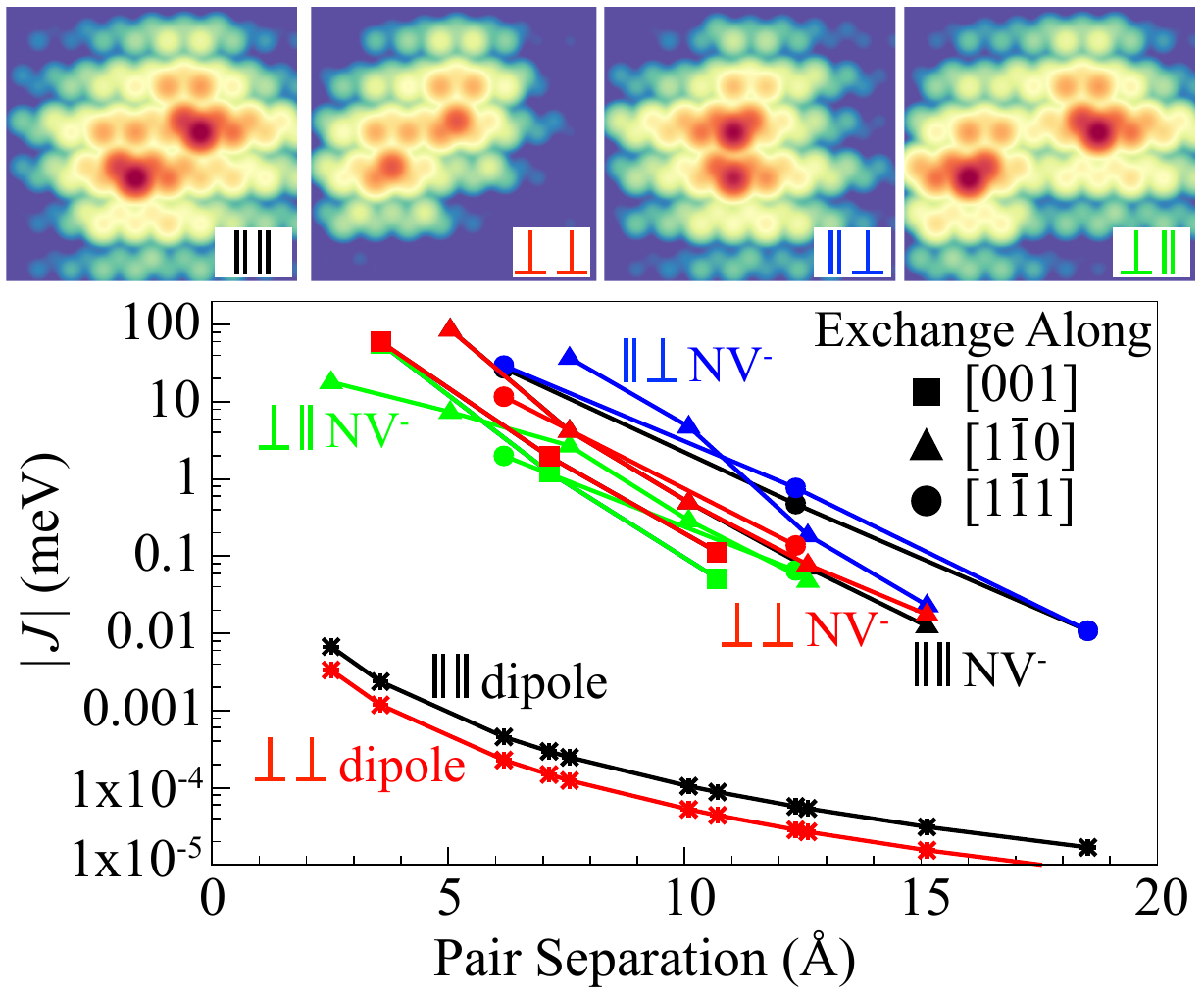}
\caption{Anisotropy of the exchange interaction for pairs of NV$^-$ centers along $[001]$, $[1\bar10]$ and $[1\bar11]$ denoted by triangles, squares and circles respectively.  The inserts are the four real space probability densities representing the different orientations of the NV$^-$ centers with respect to $[1\bar11]$ direction.  They are plotted in the (110) plane containing the centers, for two NV$^-$ centers separated by 6.17 \AA\vspace{1pt} with the same logarithmic color scale used in Fig.~\ref{wavefunction}.}
\label{NVexchange}
\end{figure} 

The NV$^-$ center exhibits an additional form of exchange interaction anisotropy, corresponding to the dependence of the exchange interaction on the relative orientation of the NV$^-$ center atoms themselves.  The vacancy and the nitrogen  can either be oriented near-parallel to $[1\bar11]$ or near-perpendicular to $[1\bar11]$.  This introduces four orientations for a pair of NV$^-$ centers, (1) both near-parallel to $[1\bar11]$, (2) both near-perpendicular to $[1\bar11]$ and (3) and (4) corresponding to types with one of the pair near-parallel and the other near-perpendicular to $[1\bar11]$, pictured in Fig.~\ref{NVexchange}.  The choice of near-parallel or near-perpendicular orientation of the NV$^-$ center has a large effect on the exchange interaction.  Due to the geometry of NV$^-$ center pairs along the $[1\bar10]$ direction, for some pairs the first nearest neighbor and in one case the second nearest neighbor exchange interactions are not presented due to overlapping impurity potentials.  As expected from the symmetry of the different pairs, in some directions there are pairs which have similar exchanges.  For example along the $[001]$ direction the exchange interactions between the near-parallel near-perpendicular (blue) and near-perpendicular near-parallel (green) overlay each other in the plot as do the values for the near-perpendicular near-perpendicular (red) and near-parallel near-parallel pairs (black).  At the largest spacings the near-parallel near-perpendicular (blue) and near-parallel near-parallel (black) pairs have the largest exchange interactions along $[1\bar11]$ in direct contrast with the transition metal pairs where the interactions along $[11\bar1]$ are in general the smallest.  Once again, beyond these pair spacings the exchange interaction is hidden by the 10\ $\mu$eV broadening included in the homogenous Green's function calculations.  

The exchange interactions between pairs of transition metal pairs of spin centers and pairs of NV$^-$ centers are comparable in magnitude.  For all the species and orientations of pairs  at the calculated separations the exchange interactions exceed the dipole-dipole interaction between two electrons regardless of dipole orientation.  Taking a linear fit to the logarithmic decrease of the exchange interaction along the $[1\bar10]$ direction, the exchange interaction between  two Mn equals the dipolar interaction at 47 \AA\vspace{1pt}; this crossover occurs at roughly 22 \AA\vspace{1pt} and 25 \AA\vspace{1pt} for the other transition metal pairs and different orientations of NV$^-$ pairs respectively.

We have constructed a detailed and accurate theoretical description of NV$^-$ and transition-metal point defect spin centers in diamond. 
The exchange interactions for pairs of transition metal spin centers are on the order, and in some cases, larger than the exchange interaction for pairs of NV$^-$ centers.  The spin 1 transition metal dopants, Cr\---Cr and Ni\---Ni, show experimentally relevant exchange interactions, in excess of the dipolar interactions between spin centers, even at 2-3~nm separations.  
Transition metal dopants in diamond offer distinct properties compared to NV$^-$ spin centers due to the inclusion of  $d$-orbitals and the resulting spin-orbit interaction that permits high-speed electrical control of spin\cite{Tang2006} and spin-sensitive optical selection rules.  Additionally, based on the exchange between a Ni and Cr dopant pair, one could envision a quantum register where information is transferred to the spin of a Ni spin center and then that information is  stored in the less accessible Cr spin. 

We acknowledge support from an AFOSR MURI.

\end{document}